\documentclass[11pt]{article}
\usepackage{tabularx}
\usepackage{subcaption}
\usepackage{multirow}
\usepackage{chngcntr}
\counterwithout{figure}{section}
\usepackage{footnote}
\makesavenoteenv{tabular}
\usepackage{microtype}
\usepackage{booktabs}
\usepackage{tikz}
\usetikzlibrary{arrows,decorations,backgrounds, calc, fit, shapes, positioning}
\usetikzlibrary{shapes.geometric}   
\usepackage{rotating}    
\usepackage{bm}
\usepackage{xcolor}
\usepackage{mwe}
\usepackage[margin=1in]{geometry}
\usepackage[english]{babel}
\usepackage{longtable}
\usepackage[T1]{fontenc}
\usepackage[utf8]{inputenc}
\usepackage{amsmath}
\usepackage{amsfonts}
\usepackage{amsthm}
\usepackage{amssymb}
\usepackage{mathtools}
\usepackage{graphicx}
\usepackage{tgtermes}
\usepackage{etoolbox}
\usepackage{enumerate}
\usepackage{algorithm, algorithmicx}
\usepackage[noend]{algpseudocode}
\usepackage[hyphens]{url}
\usepackage{tabularx}
\usepackage{hyperref}
\hypersetup{
	colorlinks=true,
	linkcolor=blue,
	filecolor=magenta,      
	urlcolor=cyan,
	pdftitle={Research Proposal - Igor Sadoune (Preliminary)},
	pdfpagemode=FullScreen,
}

\addto{\captionsenglish}{}

\DeclareMathOperator*{\argmax}{arg\,max}

\numberwithin{figure}{section}

\tikzset{
	global scale/.style={
    scale=0.6,
    every node/.style={scale=0.6}
  }
}

\title{Implementing a Hierarchical Deep Learning Approach for Simulating multilevel Auction Data}

\date{}

\begin{document}
\maketitle

\vspace{-2cm}

\begin{center}
    Igor Sadoune$^{1, 2, *}$, Marcelin Joanis$^{1, 2, \ddagger}$, Andrea Lodi$^{3, \ddagger}$\\
    \vspace{1cm}
    $^1$Department of Mathematics and Industrial Engineering, Polytechnique Montreal, Montreal, QC, Canada\\
    $^2$CIRANO, Montreal, QC, Canada\\
    $^3$Jacobs Technion-Cornell Institute, Cornell Tech and Technion - IIT, New York, NY, USA\\
    $^*$Corresponding author: igor.sadoune@polymtl.ca.\\
    $^\ddagger$These authors contributed equally to this work: marcelin.joanis@polymtl.ca, andrea.lodi@cornell.edu.\\
\end{center}

\begin{abstract}
We present a deep learning solution to address the challenges of simulating realistic synthetic first-price sealed-bid auction data. The complexities encountered in this type of auction data include high-cardinality discrete feature spaces and a multilevel structure arising from multiple bids associated with a single auction instance. Our methodology combines deep generative modeling (DGM) with an artificial learner that predicts the conditional bid distribution based on auction characteristics, contributing to advancements in simulation-based research. This approach lays the groundwork for creating realistic auction environments suitable for agent-based learning and modeling applications. Our contribution is twofold: we introduce a comprehensive methodology for simulating multilevel discrete auction data, and we underscore the potential of DGM as a powerful instrument for refining simulation techniques and fostering the development of economic models grounded in generative AI.
\end{abstract}


\vspace{0.25 cm}

\section{Introduction}
\label{sec:intro}

In this paper, we propose a hierarchical deep learning method for generating synthetic yet realistic first-price auction data. Our approach is designed to address the challenges associated with high-cardinality discrete feature spaces and multilevel structures inherent to auction data. By leveraging the capabilities of generative deep learning, our primary contribution is to provide the right methodology for crafting realistic auction simulations, which in turn can facilitate the application of agent-based modeling (ABM) and complex system approaches to the study of auction markets.

Auction markets, as a prominent example of complex systems, have gained significant attention from researchers and practitioners due to their inherent complexity and potential applications across various fields, such as finance, economics, and operations research \cite{milgrom2003}. The study of first-price auctions, in particular, is crucial for understanding the mechanisms and dynamics that govern their functioning, which are prevalent in numerous industries and applications, including online advertising \cite{edelman2007}, electricity markets \cite{hortacsu2008}, and public procurement in general \cite{uyarra2020}.

We argue that simulations, ABM, and the complex system approach, in general, may offer advantages over static statistical analysis by providing a more detailed representation of individual interactions and capturing the emergent properties of complex systems \cite{tesfatsion2006, bonabeau2002, dawid2018}. The integration of deep learning techniques, such as deep generative modeling, with the crafting of realistic auction simulations is expected to promote more robust and resilient economic systems, ultimately benefiting various fields, including computational economics \cite{athey2019c, Calpin2001, Xie2018}. 

The method developed in this paper utilizes generative adversarial networks (GANs) \cite{goodfellow2014} and variational autoencoders (VAEs) \cite{kingma2014} for the replication of the auction feature space, while employing a neural network, referred to as Bidnet, to predict the conditional bid distribution for each instance of auction from the generated feature space. Recent advancements in deep generative modeling (DGM), such as the development of the conditional tabular generative adversarial networks (CTGANs) \cite{xu2019}, have enabled the efficient handling of high-cardinality discrete distributions, which is a critical aspect for the generation of realistic auction data. Additionally, our Bidnet architecture is specifically designed to address the multilevel issue inherent to auction data, by capturing the relationships between auctions and their associated bids. This combination of DGM and Bidnet provides a comprehensive and effective approach for simulating first-price auction data, while accounting for the complexities and challenges associated with high-cardinality discrete feature spaces and multilevel structures.

\subsection{Paper Organization}
Section \ref{sec:2} delves into the mathematical background of GANs and VAEs, discussing their advantages over traditional methods for data generation. This section also explores the application of deep generative modeling in social sciences and its use in high-dimensional discrete spaces.

In Section \ref{sec:3}, we detail the process of generating synthetic multilevel auction data. This section first presents the multilevel problem and the problem formulation, followed by the solution framework. We then discuss approximating auction features joint density using GAN-based approaches and tabular variational encoding. The section continues with the training of a generator for continuous bids and concludes with the sampling of synthetic auction instances.

Section \ref{sec:4} focuses on the validation of the synthetic data generated by our proposed methods. This section evaluates the faithfulness of the synthetic auction features and the performance of BidNet in generating synthetic bids.

In Section \ref{sec:5}, we present a discussion that summarizes the study's findings, emphasizing the credibility and usefulness of our contribution. This section also discusses the limitations of our approach and highlights the superiority of CTGANs in our specific context.

Finally, Appendices A, B, and C provide additional details on data, methodology, and algorithms, respectively.

\section{Deep Generative Modeling}
\label{sec:2}

Deep generative modeling includes a variety of techniques that aim to perform density estimation using artificial neural networks as function approximators. Certain models can obtain an explicit representation of the target distribution, while others provide only an implicit, or "black box," representation of the data structure to be replicated. In this study, we focus on the latter category of models, specifically GANs \cite{goodfellow2014} and VAEs \cite{kingma2014}. The strength of these models lies in their ability to eliminate the necessity for detailed knowledge of the underlying data structure being replicated. In essence, GANs and VAEs possess the capability to generate an extensive array of synthetic data points from a limited number of empirical observations.

Adversarial learning is the core concept in GANs, which consist of two neural networks, a generator ($G$) and a discriminator ($D$), that compete in a two-player minimax game. The generator's objective is to generate synthetic samples $G(z)$, where $z$ is a random noise vector, that are similar to the true data distribution $p_{data}(x)$. The discriminator's goal is to differentiate between real samples ($x$) from the true data distribution and synthetic samples generated by the generator. The discriminator assigns a probability value $D(x)$ to each input $x$. The learning process can be described by the following objective function:

\begin{equation}
\min_G \max_D V(D, G) = \mathbb{E}{x \sim p{data}(x)}[\log D(x)] + \mathbb{E}_{z \sim p_z(z)}[\log(1 - D(G(z)))]
\end{equation}

During training, the generator and discriminator optimize this objective function in a sequential way. The generator seeks to minimize the function while the discriminator attempts to maximize it. This adversarial process leads to a convergence where the generator produces samples that the discriminator can no longer differentiate from real data samples. As the training progresses, the generator becomes increasingly skilled at producing realistic samples, while the discriminator improves its ability to distinguish between real and synthetic samples. When the equilibrium is reached, the generator generates synthetic samples that resemble the true data distribution, and the discriminator is unable to distinguish between real and generated samples, assigning a probability of $\frac{1}{2}$ to each input.

\begin{figure}[ht]
	\centering
	\caption{\footnotesize Chart of adversarial learning (left) and variational autoencoding (right).}
	\tikzstyle{neuron} = [draw, rectangle, inner sep=0.3cm, minimum size=0cm, rounded corners]
	\tikzstyle{nn} = [draw, circle, inner sep=0.3cm, minimum size=0cm, fill=gray!20]

    \begin{tikzpicture}[global scale]
	    \node[neuron] (z) {$\mathbf{z}\sim \mathcal{U}(0,1)$};
	    \node[nn] (G) [right of=z, xshift=2cm] {G};
	    \node[neuron] (fake) [right of=G, xshift=2cm] {$\tilde{\mathbf{x}}$};
	    \node[nn] (D) [below of=fake, yshift=-1cm] {D};
	    \node[neuron] (real) [left of=D, xshift=-2cm] {$\mathbf{x} \sim P_{real}$};
	    \node[neuron] (prob) [right of=D, xshift=2cm] {$\mathcal{P}$(input is real)};
	    \node[neuron] (loss) [right of=prob, xshift=2cm] {Loss};
	    \node[inner sep=0cm] (a1) [above of=G, yshift=0.5cm] {};
	    \node[inner sep=0cm] (a2) [below of=D, yshift=-0.5cm] {};
	    \node (bp1) [below of=prob, yshift=-0.3cm] {\footnotesize backpropagation}; 
	    \node (bp2) [above of=fake, xshift=1.5cm , yshift=0.3cm] {\footnotesize backpropagation}; 
	    \draw[arrows=-{latex}] (z) -- (G);
	    \draw[arrows=-{latex}] (G) -- (fake);
	    \draw[arrows=-{latex}] (fake) -- (D);
	    \draw[arrows=-{latex}] (real) -- (D);
	    \draw[arrows=-{latex}] (D) -- (prob);
	    \draw[arrows=-{latex}] (prob) -- (loss);
	    \draw[arrows=-{latex}] (loss) |- (a1.south) -- (G.north);
	    \draw[arrows=-{latex}] (loss) |- (a2.north) -- (D.south);
	    
	    \node[neuron] (in) [right of=loss, xshift=1.5cm]{$\mathbf{x} \sim P_{real}$};
	    \node[nn] (E) [right of=in, xshift=1cm] {E};
	    \node[neuron] (mu) [right of=E, xshift=1cm, yshift=0.5cm] {$\mu$};
	    \node[neuron] (sigma) [right of=E, xshift=1cm, yshift=-0.5cm] {$\sigma$};
	    \node[neuron] (sample) [right of=mu, xshift=1.5cm, yshift=-0.5cm] {$\mathbf{z} \sim \mathcal{N}(\mu, \sigma)$};
	    \node[nn] (D2) [right of=sample, xshift=2cm] {D};
	    \node[neuron] (out) [right of=D2, xshift=1cm] {$\tilde{\mathbf{x}}$};
	    \node[neuron] (loss2) [above of=mu, yshift=1cm] {loss};
	    \node[inner sep=0cm] (a12) [above of=in, yshift=1.5cm] {};
	    \node[inner sep=0cm] (a22) [above of=out, yshift=1.5cm] {};
	    \scoped[on background layer]
	    \node (code) [draw, thick, rounded corners, 
	    inner sep=.6cm, label=below:Code, 
	    fit=(mu) (sigma)] {};
	    \draw[arrows=-{latex}] (in) -- (E);
	    \draw[arrows=-{latex}] (E) -- (mu);
	    \draw[arrows=-{latex}] (E) -- (sigma);
	    \draw[arrows=-{latex}] (mu) -- (sample);
	    \draw[arrows=-{latex}] (sigma) -- (sample);
	    \draw[arrows=-{latex}] (sample) -- (D2);
	    \draw[arrows=-{latex}] (D2) -- (out);
	    \draw[arrows=-{latex}] (code) -- (loss2);
	    \draw[arrows=-{latex}] (in.north) -| (a12) -- (loss2.west);
	    \draw[arrows=-{latex}] (out.north) -| (a22) -- (loss2.east);
	    \draw[arrows=-{latex}] (loss2) -- (E.north);
	    \draw[arrows=-{latex}] (loss2) -- (D2.north);

    \end{tikzpicture}

	\label{fig:ganvae}
\end{figure}
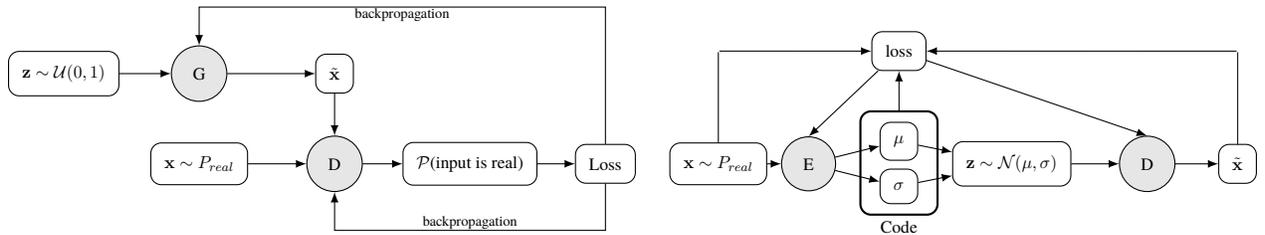

VAEs are a type of generative model that leverages variational inference to learn a probabilistic mapping between the data and a latent space. The key idea behind VAEs is to introduce a probabilistic encoder $q_\phi(z|x)$, that approximates the true posterior distribution $p_\theta(z|x)$, and a probabilistic decoder $p_\theta(x|z)$, that models the data distribution conditioned on the latent variable $z$. Here, $\phi$ and $\theta$ represent the parameters of the encoder and decoder neural networks, respectively. Variational inference is used to optimize the evidence lower bound (ELBO) on the log-likelihood of the data, i.e.,

\begin{equation}
\log p_\theta(x) \geq \mathbb{E}{q\phi(z|x)}[\log p_\theta(x|z)] - D_{KL}(q_\phi(z|x) || p_\theta(z)) = \mathcal{L}(\theta, \phi; x).
\end{equation}

The ELBO consists of two terms: the reconstruction term, $\mathbb{E}{q\phi(z|x)}[\log p_\theta(x|z)]$, which measures the ability of the model to reconstruct the data given the latent variable, and the regularization term, $D_{KL}(q_\phi(z|x) || p_\theta(z))$, which measures the divergence between the approximate posterior and the prior distribution over the latent space. During training, the VAE jointly optimizes the encoder and decoder parameters to maximize the ELBO with respect to the model parameters $\theta$ and $\phi$. This process encourages the learned latent space to have a meaningful structure, enabling efficient generation of new samples by sampling from the prior $p_\theta(z)$ and decoding them using the decoder $p_\theta(x|z)$.

\subsection{Traditional Methods for Data Generation}

Market simulations and simulation crafting have been widely used across various disciplines to study complex systems, replicate real-world dynamics, and develop policies and strategies. The use of market simulations can be traced back to the early works of economists and computer scientists \cite{arthur1999}. Agent-based modeling has emerged as a prominent approach to studying markets and economic systems, allowing researchers to capture the interactions between heterogeneous agents and their influence on market outcomes \cite{tesfatsion2006}. A large body of literature has explored different aspects of market simulations, from designing auction mechanisms \cite{decarolis2017} to investigating the dynamics of financial markets \cite{marti2020}. 

Historically, traditional methods for simulation crafting, such as Monte Carlo sampling, bootstrapping, cellular automata, and system dynamics modeling, have been widely used in various applications \cite{rubinstein2008, tesfatsion2006, dvison1986}. Monte Carlo sampling generates random samples from a given probability distribution, cellular automata simulates spatial and temporal dynamics through discrete cells evolving based on predefined rules, and system dynamics modeling uses differential equations to describe relationships among system components. 

However, these traditional methods exhibit limitations when applied to complex systems with high-cardinality discrete spaces or multilevel structures, such as auction markets. These limitations include difficulties in accurately representing high-dimensional probability distributions, modeling intricate dependencies between variables, accommodating nonlinear dynamics and non-stationary behaviors, and integrating domain-specific knowledge or constraints \cite{elsawah2020}. Furthermore, traditional simulation methods often require substantial computational resources and are constrained by their assumptions and rules \cite{bonabeau2002}.

The advent of machine learning and artificial intelligence has significantly impacted simulation crafting, paving the way for advanced techniques such as DGM. Deep generative models including GANs  and VAEs, address the limitations of traditional techniques by learning and representing high-dimensional probability distributions, modeling complex dependencies between variables, and accommodating nonlinearities and non-stationarities in complex systems \cite{Chen2016}. These innovative methods have been applied to various domains, such as finance \cite{axtell2021}. Although GANs have been recognized as having potential for creating more realistic market simulations and fostering the development of more effective policies and strategies \cite{Athey2019d}, the use of GANs in economics remains limited, with only a small number of applications currently found in the field.

Leveraging unsupervised learning and deep neural networks, DGM methods identify and represent the underlying data structure in a flexible and scalable manner. Additionally, DGM techniques can model complex dependencies and correlations between variables, providing a more accurate representation of the relationships present in the data. DGM techniques suitable for handling the nonlinearities and non-stationarities that are inherent to complex systems. Due to their deep architectures and non-linear activation functions, deep generative models can capture complex, multi-scale structures in data, which may be challenging for traditional methods to accurately represent. Moreover, DGM approaches can be adapted to incorporate domain-specific knowledge and constraints, further enhancing the realism and validity of the generated synthetic data.

\subsection{DGM in Social Sciences}
\label{subsec:dgmsocialsciences}
Although deep generative models are still underused in social sciences, they can potentially provide substantial improvements in any application relying on qualitative data. 

They have emerged as a powerful tool for addressing low degrees of freedom. Specifically, DGMs amplify observations and enhance weak signals within categorical configurations, enabling the generation of synthetic data that closely approximates the characteristics of real-world data. This technology has found applications in a diverse range of domains, such as credit card datasets \cite{Ba2019} and medical data recovery \cite{Jackson2019}.

In causal effect research, DGMs play a pivotal role by enabling counterfactual analysis using propensity scores and facilitating the estimation of unseen or partially unseen distributions through GANs \cite{Yoon2018} and VAEs \cite{Louizos2017b, cai2019a}. Additionally, DGMs have been shown to benefit privacy-sensitive applications involving medical or financial data, leading to specialized literature in this area \cite{Cai2021}.

Moreover, DGMs have contributed to the reinvigoration of agent-based modeling in economics by generating synthetic data for creating realistic artificial environments and maintaining simulation realism \cite{axtell2021}. For example, DGM-based travel behavior simulations have employed restricted Boltzmann machines \cite{Wong2020}, while GAN-based financial correlation matrices and time-series sampling have been used for simulating financial systems \cite{marti2020, Takahashi2019}.

\subsection{GANs for High-Dimensional Discrete Spaces}
Despite the progress in market simulations and simulation crafting, several challenges remain. In our case, generating high-cardinality discrete data and multilevel data structures is still a complex task, as highlighted by \cite{xu2019}. Furthermore, incorporating the dynamics of real-world markets and their ever-evolving nature into simulation models requires continuous research and development. Various strategies have been developed to enhance the stability and performance of GANs, particularly focusing on innovations that address the challenges of handling discrete inputs. 

For instance, the Wasserstein GAN (WGAN) \cite{Arjovsky2017, Gulrajani2017} and the least-square GAN (LSGAN) \cite{Mao2017} employ a critic rather than a traditional discriminator. The critic predicts the distance between a given point and the decision boundary separating real and fake samples, thereby improving the signal quality for the generator during training and leading to better convergence properties. WGAN's popularity stems from its Earth Mover (EM) or Wasserstein-I loss, which measures the distance between real and fake sample distributions, thus promoting stable and robust training \cite{Arjovsky2017}.

The boundary-seeking GAN (BGAN) \cite{Hjelm2019} introduces a modified training process, where the generator uses a policy gradient that accommodates both discrete and continuous inputs. This approach results in smoother training and mitigates issues related to mode collapse, a common problem in GANs. 

Meanwhile, the conditional tabular GAN (CTGAN) \cite{xu2019} is a framework specifically designed to address data imbalances and discrete inputs in tabular data. Functioning as a meta-algorithm, CTGAN is compatible with various loss functions, network topologies, or training processes. The core concept of CTGAN, training by sampling, augments the generator's input space with a conditional vector that encodes the selection, thereby enhancing the model's ability to capture the underlying structure of discrete data and generate realistic synthetic samples.

\section{Generating Synthetic multilevel Auction Data}
\label{sec:3}
In this study, we aim to create realistic synthetic auctions using a novel method that combines two distinct functions, each handling a specific aspect of the auction data. The first function generates artificial auction characteristics, capturing the features of the contracts being offered. The second function,later introduced as BidNet, approximates the distribution of bids given these auction features, providing an aggregated representation of the bidding firms. By breaking down the process into these two sequential functions, our method addresses technical challenges while maintaining the overall structure of the auction. We validate our approach by training a predictor using the synthetic data and evaluating its performance on real-world data. This method effectively separates the generation of discrete and continuous data types and simplifies the complex, multilevel auction structure, making it suitable for generating realistic market simulations.

We have chosen the context of public procurement to demonstrate our ability to simulate realistic first-price auctions effectively. To this end, we rely on the data provided by the "Système électronique d'appel d'offre" (SEAO), which encompasses 117,249 contracts offered in public market auctions within the Canadian province of Quebec over a ten-year period (2010-2020). In this scenario, the high-cardinality discrete feature space represents various auction attributes, such as the type of contract or the designated delivery region. The multilevel structure of the auction data emerges from the multiple firms participating in the bidding process for each auction. Utilizing this comprehensive dataset allows us to develop and refine our simulation methodology, thereby showcasing its potential for generating accurate and realistic representations of first-price auctions in the domain of public procurement.

\subsection{Data Specification}
Table \ref{tbl:1} summarizes the structure of the data after cleaning. We differentiate between \textit{multiclass} and  \textit{multilabel} variables. A multicalss variable is categorical, but each instance belongs to exactly one out of a finite set of classes. On the other hand, a \textit{multilabel} variable represents cases where an observation can simultaneously have multiple values. For instance, in our scenario, an auction could be either municipal or provincial, making the variable \textit{municipality} a binary multiclass variable. It encodes whether an auction is issued at the municipal level ($1$) or not ($0$). Conversely, the variable \textit{firms} can take varying numbers of values per auction, as multiple firms may participate in an auction. Therefore, the variable \textit{firms} is multilabel. This implies that the cardinality (the number of possible states) of such a variable increases exponentially with the number of potential values.

\begin{table}[ht]
\caption{Structure of the data.}
	\centering
	\begin{tabular}{l c c}
	\toprule
	\textbf{Variables} & \textbf{Type} & \textbf{Cardinality}\\
	\hline
      \textit{public contractor}     & multiclass & 1,451  \\
      \textit{municipality}          & multiclass & binary  \\
      \textit{sector}                & multiclass & 3  \\
      \textit{subsector}             & multiclass & 53  \\
      \textit{location}              & multiclass & 98  \\
      \textit{unspsc}                & multiclass & 1,379  \\
      \textit{number of bidders}     & multiclass & 24 \\
      \textit{post-auction expenses} & multiclass & binary \\
      \textit{firms}                 & multilabel & 40,659 \\
      \textit{bids}                  & continuous &    -   \\
	\bottomrule
	\end{tabular}
\label{tbl:1}
\end{table}

We can see in Table \ref{tbl:1} that the multilabel variable \textit{firms} is relatively large compared to the dataset. This indicates that a significant number of firms have participated in the auctions included in our data, which highlights the multilevel issue described in Section \ref{subsec:multilevel}.

The data comprises 9 categorical auction features (excluding \textit{firms}), spans across 3 sectors (construction, supply, and services), and includes a total of 117,249 auctions and 448,935 bids. The selected set of variables shown in Table \ref{tbl:1} represents a subset of relevant, high-quality, and informative variables from the columns provided by the SEAO.

A textual description of the variables used in the paper, along with a guide to access, download, and process the raw data from the SEAO, including a link to the official SEAO PDF document that describes all the features, can be found in Appendix A. Additional details about the raw data, including cleaning and preprocessing procedures, are available in the GitHub repository associated with this manuscript.

\subsection{The Multilevel Problem and Firm Representation}
\label{subsec:multilevel}
Our dataset highlights the multilevel nature of first-price sealed-bid auctions, where each auction is linked to a varying number of bids and firms, creating a complex set of auction features alongside a combinatorial multilabel subspace of bids and firm characteristics. However, access to detailed and high-quality firm features is often limited in public procurement data, such as in the SEAO datasets. While it is theoretically possible to infer such features (e.g., cost functions, capacities), challenges such as data scarcity—where over 90\% of firms are infrequently represented in auctions—and market dynamics, including mergers, name changes, or new entries, complicate tracking firm activities over time. Additionally, the exponential growth in complexity with each new firm added and the presence of diverse market sectors further hinder a direct modeling approach.

The medical GANs (MedGAN), developed in \cite{Choi2017} and \cite{Jackson2019}, offers a potential framework by approximating the joint probability distribution of auction features with GANs, then applying autoencoders to learn the multilabel subspaces (bids and firms' features). However, the high number of firms (see Table \ref{tbl:1}) complicates simultaneous optimization of GANs and autoencoders, suggesting that a direct application of MedGAN to our context may not be feasible without significant data reduction, which would undermine our objectives.

Given these challenges, our study proposes a hierarchical and modular deep learning solution akin to MedGAN's approach. We employ deep generative modeling to sample from the joint distribution of auction features and introduce BidNet, a supervised learning model that implicitly captures the intricate relationship between auction characteristics and associated combinatorial subspace. This method, focusing on a high-level representation of firms, simplifies the complexity of the multilevel problem, but is advanced enough to enable more realistic auction simulations, marking a significant step forward in addressing the challenges of replicating multilevel auction data.

\subsection{Problem Formulation and Solution Framework}
An instance of public procurement auction is labeled $a\in\{1,\dots,N\}$, where $N$ is the number of examples. Let $C$ be a collection of multiclass variables representing the features of the contracts being offered, and let $nb \in C$ be the number of bidders in an auction. We adopt a sparse one-hot encoded form of our discrete space, meaning that each $c\in C$ is represented by a binary vector such that $\sum_i c(i) = 1$, and $\mathbf{c}$ is the aggregated vector of multiclass variables such that $\sum_j \mathbf{c}(j) = |C|$. Introducing the variable \textit{firms}, we represent it by the binary variable $\mathbf{f}$, which encodes the presence or absence of firms bidding in each auction, such that $\sum_i \mathbf{f}(i) = nb^a$ for each instance $a$. Thus, an auction $a$ is given by the set $\mathbf{x} = \{ \mathbf{c}^a, \mathbf{f}^a, \mathbf{b}^a \}$, where $\mathbf{b}$ is an array of continuous bids. It follows that the number of elements in $\mathbf{b}^a$ equals $nb^a$. Let $a^*$ be a hypothetical auction with only one feature ($|C|=1$), and two bidders among four firms composing the market. Then, $nb^{a^*}=2$, and $\mathbf{x}^{a^*}$ could be (depending on which firms are actually bidding) $\{ [1, 0], [1,1,0,0], [b_1, b_2] \}$, where $b_1, b_2 > 0$. The problem of density estimation in our case is to approximate the resulting joint distribution $p(\mathbf{x})$. Since we wish to generate new samples, our problem is to optimize the set of parameters $\psi$ for the mapping $M: \mathbf{x} \xrightarrow{p(\mathbf{x};\psi)}\tilde{\mathbf{x}}$, where $\tilde{\mathbf{x}}$ is a fake but realistic synthetic auction, i.e., it is impossible to say if $\tilde{\mathbf{x}}\sim p(\mathbf{x})$ or $\tilde{\mathbf{x}}\sim p(\cdot)$, where $p(\cdot)$ is an arbitrary probability function.

We can decompose the mapping $M$ into two independent and sequential functions, specifically by considering the functions $A:\mathbf{z} \xrightarrow{p(\mathbf{c|\mathbf{z}}; \alpha)}\tilde{\mathbf{c}}$ and $B:\mathbf{c}\xrightarrow{p(b|\mathbf{c}; \beta)}\hat{\theta}$. The function $A$ generates synthetic samples of auction characteristics, $\tilde{\mathbf{c}}$, from the noise input $\mathbf{z}$, by approximating $p(\mathbf{c}|\mathbf{z}; \alpha)$ using the set of parameters $\alpha$. Utilizing the latent signal $\mathbf{z}$ is advantageous because synthetic samples can be generated from scratch, with real data required only for training the approximator.

The function $B$ approximates the conditional distributions of the bids given auction features, thus providing an aggregated representation of the firms. The vector $\hat{\theta}$ consists of the estimated parameters for the conditionals and depends on the probability function employed to describe the distribution of bids. Consequently, the bid generator $B$ produces estimates that serve as arguments for a random generator, from which $nb$ bids are sampled, i.e., $\hat{b} \sim \mathcal{P}(\hat{\theta})$.

While the specific functional forms for $A$ and $B$ have not yet been determined, we can already observe how the general structure coherently articulates. Once both models are trained with respect to $\alpha$ and $\beta$, we can generate synthetic but realistic vectors of auction features, $\tilde{\mathbf{c}}$, using function $A$. These vectors are then utilized as input for function $B$, reconstructing the space ${\mathbf{c}, \mathbf{b}}$. Consequently, the generation of each data type is separated between $A$ (discrete) and $B$ (continuous), effectively flattening the multilevel auction structure, as $B$ represents all firms simultaneously. At this stage, the problem initially formalized by the mapping $M$ has been divided into functions $A$ and $B$, or equivalently, we have defined $\psi = (\alpha, \beta)$.

The proposed pipeline does not explicitly incorporate the multilabel variable \textit{firms}, as its cardinality is too large to feasibly preserve its original form and interpretability. Furthermore, we find it counterproductive to use a smaller continuous code representing \textit{firms} if the original space cannot be retrieved subsequently. It is also worth noting that retaining only a subset of the data to reduce cardinality would undermine the objective.

\begin{figure}[!ht]
\tikzstyle{neuron} = [draw, rectangle, inner sep=0.3cm, minimum size=0cm, rounded corners]
\tikzstyle{lbl} = [draw, circle, inner sep=0.05cm, minimum size=0cm, fill=gray!20]
    \caption{\footnotesize The graph illustrates the sequential structure of our meta-algorithm. The approximators $A$ and $B$ are trained, with regard to their respective set of parameters  $\alpha$ and $\beta$, on real examples. Then, the solidified forms  $A^*$ and $B^*$ are used to generate the synthetic features $\tilde{\mathbf{c}}$ and bids $\tilde{\mathbf{b}}$.} 
	\centering
    \begin{tikzpicture}
    \node[neuron] (alpha) {$\alpha:\xrightarrow{A(\mathbf{c}_{train})}\alpha^*$};
    \node[neuron] (beta) [below of=alpha, yshift=-0.5cm] {$\beta:\xrightarrow{B(\mathbf{c}_{train}, \mathbf{b}_{train})}\beta^*$};
    \node[neuron, fill=gray!20] (ctilde) [right of=alpha, xshift=4cm] {$\overset{\sim}{\mathbf{c}} \sim A_{\alpha^*}(\mathbf{z})$};
    \node[neuron] (z) [above of=ctilde, yshift=0.5cm] {$\mathbf{z}\sim \mathcal{N}(0,1)$};
    \node[neuron] (thetatilde) [right of=beta, xshift=4cm] {$\tilde{\theta} \sim B_{\beta^*}(\tilde{\mathbf{c}})$};
    \node[neuron, fill=gray!20] (btilde) [below of=thetatilde, yshift=-0.5cm] {$\tilde{b} \sim \mathcal{N}(\tilde{\theta})$};
    \node[neuron] (gamma) [right of=ctilde, xshift=3cm] {$\gamma:\xrightarrow{F(\tilde{\mathbf{c}})}\gamma^*$};
    \node[neuron] (incepscore) [right of=gamma, xshift=2.5cm] {$score\big(F_{\gamma^*}(\mathbf{c}_{test})\big)$};
    \node[neuron] (bidnetscore) [right of=thetatilde, xshift=3.5cm] {$score\big(B_{\beta^*}(\mathbf{c}_{test})\big)$};
    
    \draw[arrows=-{latex}] (alpha) -- (ctilde);
    \draw[arrows=-{latex}] (beta) -- (thetatilde);
    \draw[arrows=-{latex}] (z) -- (ctilde);
    \draw[arrows=-{latex}] (ctilde) -- (thetatilde);
    \draw[arrows=-{latex}] (thetatilde) -- (bidnetscore);
    \draw[arrows=-{latex}] (ctilde) -- (gamma);
    \draw[arrows=-{latex}] (gamma) -- (incepscore);
    \draw[arrows=-{latex}] (thetatilde) -- (btilde);

    \scoped[on background layer]
    \node (Training) [draw, thick, rounded corners, 
    inner sep=.3cm, label=above:Training, 
    fit=(alpha) (beta)] {};
    \node (Sampling) [draw, thick, rounded corners, 
    inner sep=.3cm, label=above:Sampling, 
    fit=(z) (btilde)] {};
    \node (Validation) [draw, thick, rounded corners, 
    inner sep=.3cm, label=above:Validation, 
    fit=(gamma) (bidnetscore) (incepscore)] {};
    \end{tikzpicture}
\label{fig:architecture}
\end{figure}
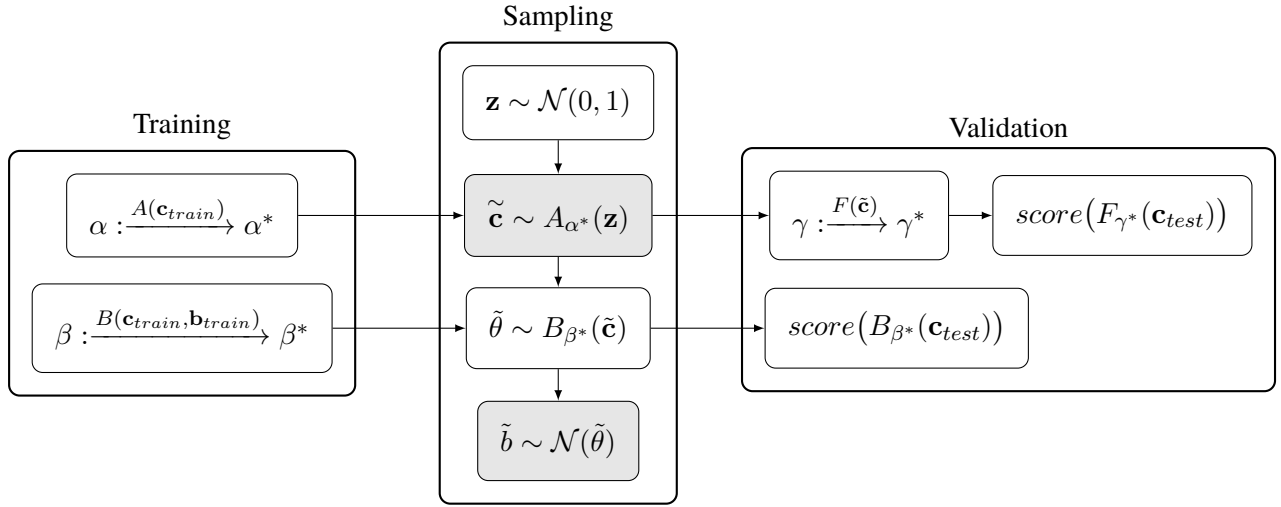

For validation, we employ the inception scoring method, which involves training a predictor $F$ using the newly sampled synthetic data and evaluating its performance on the real data. The bid generator is subsequently assessed separately with real examples that were not seen during training. Figure \ref{fig:architecture} offers a visual representation of the comprehensive system we have outlined.

\subsection{Approximating Auction Features Joint Density}
\label{subsec:ta}
This section discusses the details regarding the training of two synthesizers based on GANs and VAEs for the mapping $A:\mathbf{z} \xrightarrow{p(\mathbf{c|\mathbf{z}}; \alpha)}\tilde{\mathbf{c}}$. It is essential to note that they are not complementary but competitors. It is a priori not possible to predict which one will perform best, so both are tried. Furthermore, they both deserve to be discussed in the context of our endeavor.  

\subsubsection{GAN-based Approach}
\label{subsubsec:gans}
Our GAN-based algorithm consists of a generator $G$ and a critic $C$ that we optimize with respect to the Wasserstein loss. The "training-by-sampling" principle is applied. In this context, $G$ approximates $p(\mathbf{c}|\mathbf{z}, cond)$, where $cond$ is a binary conditional vector of size $|C|$ that sums to one. Consequently, the loss function must be augmented with a cross-entropy penalty term that enforces the sampling of a synthetic data point admitting $c_{i^{*}}=1$, where $i^*$ is the selected state of the chosen variable $c$. The resulting objective function is expressed as follows:

\begin{equation}
\min_{G}\max_{C}E\bigg[ C(G(\mathbf{z})) - C(\mathbf{c}) \bigg] + CE(\tilde{c}_{i^{*}}, cond).
\label{eq:ctwganloss}
\end{equation}

The training procedure surrounding Equation (\ref{eq:ctwganloss}) is detailed by Algorithm \ref{alg:ctwgan} in Appendix C while Figure \ref{fig:ctwgan} provides a graphical account of the algorithmic architecture of the model. 

\begin{figure}[ht!]
	\caption{\footnotesize Neural representation of the generator and the critic used to synthesize auction characteristics. Note that $W$ and $s$ are respectively weight matrices and biases, while $h$ are outputs from hidden blocks. The dimensions of $h$, and therefore of $W$ and $s$, depend on the width parameters for a given network. The critic $C$ outputs a scalar while $G$ outputs a continuous array of size $N_x$ (the number of continuous variables) and $N_c$ one-hot arrays (one for each discrete multiclass variable).}
	\centering
	\tikzstyle{neuron} = [draw, circle, inner sep=0.3cm, minimum size=0cm]
	\tikzstyle{output} = [draw, rectangle, inner sep=0.3cm, minimum size=0cm, fill=gray!20]
	
	\begin{subfigure}[b]{0.75\textwidth}
		\centering
		\begin{tikzpicture}[global scale]
				\node[neuron] (i1) {$\mathbf{z}$};
				\node[neuron, inner sep=0.1cm] (i2) [below of=i1, yshift=-0.5cm]{$cond$};
			    \scoped[on background layer]
			    \node (i) [draw, thick, rounded corners, 
			    inner sep=.8cm, label=above:input, 
			    fit=(i1) (i2)] {};

				\node[neuron] (n1) [right of=i1, xshift=2cm, yshift=0.25cm]{};
				\node[neuron] (n2) [below of=n1, yshift=-0.5cm]{};
				\node[neuron] (n3) [right of=n1, xshift=0.5cm, yshift=0.5cm]{};
				\node[neuron] (n4) [below of=n3, yshift=-0.5cm]{};
				\node[neuron] (n5) [below of=n4, yshift=-0.5cm]{};
				\foreach \x in {1,...,2}
					\foreach \y in {3,...,5}
						\draw (n\x) -- (n\y);
			    \scoped[on background layer]
			    \node (h) [draw, thick, rounded corners, 
			    inner sep=.9cm, label=above:hidden block, 
			    fit=(n1) (n3) (n5)] {};

			    \node[output] (o2) [right of=n4, xshift=5cm] {$\tilde{c}_j \sim softmax(W_2h_2+s_2), \quad \forall j \in \{1,\dots, N_c\}$};
			
			\draw[arrows=-{latex}] (i) -- (h);
			\draw[arrows=-{latex}] (h) -- (o2.west);

		\end{tikzpicture}

		\caption{Generator}
	\end{subfigure}
	
	\vspace{0.5cm}
	
	\begin{subfigure}[b]{0.75\textwidth}
		\centering

		\tikzstyle{input} = [draw, circle, inner sep=0.005cm, minimum size=0cm]
		\begin{tikzpicture}[global scale]
				\node[neuron] (i1) {$\mathbf{c}$};
				\node[neuron] (i3) [below of=i1, yshift=-0.5cm]{$\mathbf{x}$};

			    \scoped[on background layer]
			    \node (i) [draw, thick, rounded corners, 
			    inner sep=0.8cm, label=above:input, 
			    fit=(i1) (i3)] {};

				\node[neuron] (n1) [right of=i1, xshift=3cm, yshift=0.25cm] {};
				\node[neuron] (n2) [below of=n1, yshift=-0.5cm]{};
				\node[neuron] (n3) [right of=n1, xshift=0.5cm, yshift=0.5cm]{};
				\node[neuron] (n4) [below of=n3, yshift=-0.5cm]{};
				\node[neuron] (n5) [below of=n4, yshift=-0.5cm]{};;
				\foreach \x in {1,...,2}
					\foreach \y in {3,...,5}
						\draw (n\x) -- (n\y);
			    \scoped[on background layer]
			    \node (h) [draw, thick, rounded corners, 
			    inner sep=.9cm, label=above:hidden block, 
			    fit=(n1) (n3) (n5)] {};

			    \node[output] (o) [right of=n4, xshift=3cm] {score $= Wh+s$};
			
			\draw[arrows=-{latex}] (i) -- (h);
			\draw[arrows=-{latex}] (h) -- (o.west);

		\end{tikzpicture}
	\caption{Critic}
	\end{subfigure}
	
	\label{fig:ctwgan}
\end{figure}
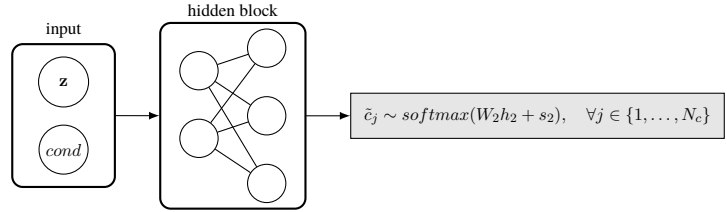

We use the PacGAN configuration \cite{Lin2018} in order to prevent mode collapse, a common problem in GANs that refers to when the model fails to capture the diversity of the underlying data distribution and instead produces a limited set of similar or identical output samples \cite{Salimans2016}. The PacGAN configuration expands the input space of the critic with multiple stacks of the original input space.
 
Additionally, we implement a gradient penalty technique, also known as WGAN-GP, instead of the traditional "weight clipping" method of imposing a Lipschitz constraint on the critic. This approach stabilizes the training process and encourages the critic to learn the correct gradients by imposing a penalty on the norm of the critic's gradient with respect to its inputs. \cite{Gulrajani2017}.

In the original GAN framework, the training process alternates between steps for the discriminator and the generator. The parameter $k$, a positive integer, determines how many times the discriminator is trained for each time the generator is trained. However, in the context of Wasserstein GANs, it is typically recommended to train the generator as frequently as the critic, essentially setting $k=1$. This is to prevent the critic from overpowering the generator too quickly. We recall that in the context of wasserstein GANs, the critic replaces the discriminator of the original framework, and predicts the distance between a given point and the decision boundary separating real and fake samples (instead of predicting the probability that the input is real).  

Finally, we use the adaptive moment estimation (Adam) optimizer, an extension of the stochastic gradient descent that improves performance by adapting the learning rate of each weight parameter based on its first and second moments \cite{kingma2015}.

\subsubsection{Tabular Variational Autoencoding}
\label{subsec:tvae}
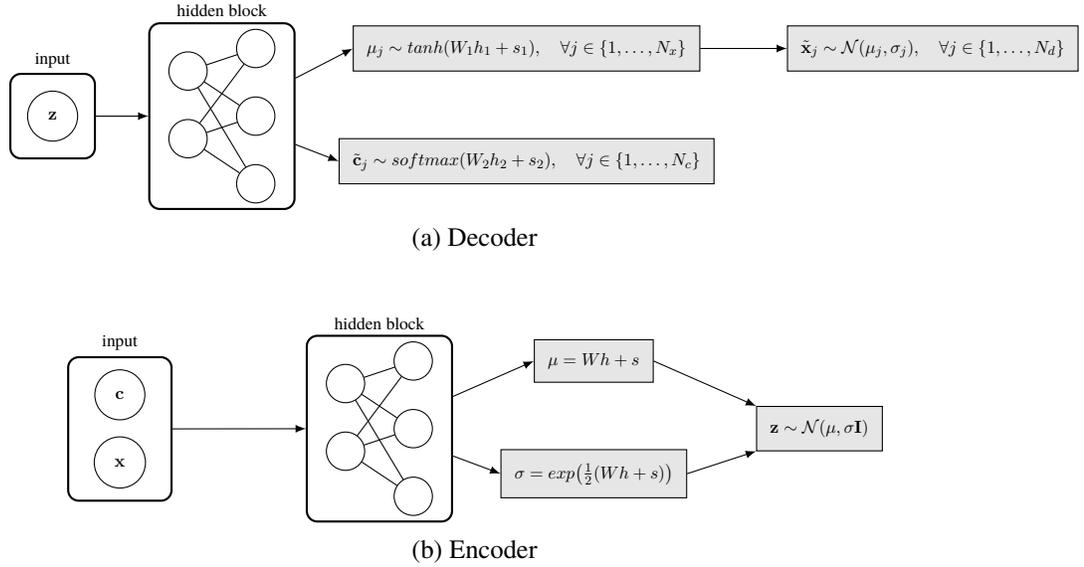
\begin{figure}[ht!]
	\caption{\footnotesize Neural representation of the encoder and decoder composing the tabular VAE used to synthesize auction characteristics. Similarly to Figure \ref{fig:ctwgan}, $W$ and $s$ are respectively weight matrices and biases, while $h$ are outputs from hidden blocks. The $\sigma_j$ are parameters of the decoder.}
	\centering
	\tikzstyle{neuron} = [draw, circle, inner sep=0.3cm, minimum size=0cm]
	\tikzstyle{output} = [draw, rectangle, inner sep=0.3cm, minimum size=0cm, fill=gray!20]
	
	\begin{subfigure}[b]{0.75\textwidth}
		\centering 
		
		\begin{tikzpicture}[global scale]
				\node[neuron] (i1) {$\mathbf{z}$};
			    \scoped[on background layer]
			    \node (i) [draw, thick, rounded corners, 
			    inner sep=0.6cm, label=above:input, 
			    fit=(i1)] {};
			
				\node[neuron] (n1) [right of=i1, xshift=2cm, yshift=1cm] {};
				\node[neuron] (n2) [below of=n1, yshift=-0.5cm]{};
				\node[neuron] (n3) [right of=n1, xshift=0.5cm, yshift=0.5cm]{};
				\node[neuron] (n4) [below of=n3, yshift=-0.5cm]{};
				\node[neuron] (n5) [below of=n4, yshift=-0.5cm]{};;
				\foreach \x in {1,...,2}
					\foreach \y in {3,...,5}
						\draw (n\x) -- (n\y);
			    \scoped[on background layer]
			    \node (h) [draw, thick, rounded corners, 
			    inner sep=.9cm, label=above:hidden block, 
			    fit=(n1) (n3) (n5)] {};

			    \node[output] (o1) [right of=n3, xshift=5cm] {$\mu_j \sim tanh(W_1h_1+s_1), \quad \forall j \in \{1,\dots, N_x\}$};
			    \node[output] (o2) [below of=o1, yshift=-1.5cm] {$\tilde{\mathbf{c}}_j \sim softmax(W_2h_2+s_2), \quad \forall j \in \{1,\dots, N_c\}$};
			    \node[output] (o3) [right of=o1, xshift=8cm] {$\tilde{\mathbf{x}}_j \sim \mathcal{N}(\mu_j, \sigma_j), \quad \forall j \in \{1,\dots, N_d\}$};
			
			\draw[arrows=-{latex}] (i) -- (h);
			\draw[arrows=-{latex}] (h) -- (o1.west);
			\draw[arrows=-{latex}] (h) -- (o2.west);
			\draw[arrows=-{latex}] (o1.east) -- (o3.west);

		\end{tikzpicture}
		\caption{Decoder}
		\end{subfigure}

\vspace{0.5cm}

	\begin{subfigure}[b]{0.75\textwidth}
		\centering
		
		\tikzstyle{input} = [draw, circle, inner sep=0.005cm, minimum size=0cm]
		\begin{tikzpicture}[global scale]
				\node[neuron] (i1) {$\mathbf{c}$};
                \node[neuron] (i2) [below of=i1, yshift=-0.5cm]{$\mathbf{x}$};
                
			    \scoped[on background layer]
			    \node (i) [draw, thick, rounded corners, 
			    inner sep=0.8cm, label=above:input, 
			    fit=(i1) (i2)] {};

				\node[neuron] (n1) [right of=i1, xshift=4cm, yshift=0.25cm] {};
				\node[neuron] (n2) [below of=n1, yshift=-0.5cm]{};
				\node[neuron] (n3) [right of=n1, xshift=0.5cm, yshift=0.5cm]{};
				\node[neuron] (n4) [below of=n3, yshift=-0.5cm]{};
				\node[neuron] (n5) [below of=n4, yshift=-0.5cm]{};
				\foreach \x in {1,...,2}
					\foreach \y in {3,...,5}
						\draw (n\x) -- (n\y);
			    \scoped[on background layer]
			    \node (h) [draw, thick, rounded corners, 
			    inner sep=.9cm, label=above:hidden block, 
			    fit=(n1) (n3) (n5)] {};

			    \node[output] (o1) [right of=n3, xshift=3cm] { $\mu = Wh+s$};
			    \node[output] (o2) [below of=o1, yshift=-1.5cm] { $\sigma = exp\big(\frac{1}{2}(Wh+s)\big)$};
			    \node[output] (o3) [right of=n4, xshift=8cm] { $\mathbf{z} \sim \mathcal{N}(\mu, \sigma \mathbf{I})$};
			\draw[arrows=-{latex}] (i) -- (h);
			\draw[arrows=-{latex}] (h) -- (o1.west);
            \draw[arrows=-{latex}] (h) -- (o2.west);
			\draw[arrows=-{latex}] (o1.east) -- (o3.north west);
            \draw[arrows=-{latex}] (o2.east) -- (o3.south west);
		\end{tikzpicture}

		\caption{Encoder}
	\end{subfigure}
	
	\label{fig:tvae}
\end{figure}

In the variational autoencoding framework, the encoder models $p(\mathbf{z}|\mathbf{c}, \mathbf{x})$, while the decoder replicates the original space by approximating $p(\mathbf{c}, \mathbf{x}|\mathbf{z})$. "Training-by-sampling" is thus neither necessary nor possible since the conditional vector would be encoded in a continuous layer. It means that only the decoder needs to be modified to generate tabular data. In addition, the reconstruction loss needs to be augmented with a cross-entropy term to ensure the integrity of the discrete structure. Algorithm \ref{alg:tvae} in Appendix C details the training procedure of the tabular variational autoencoder (TVAE), and Figure \ref{fig:tvae} provides its pictorial representation.

\subsection{Training a Generator of Continuous Bids}
\label{subsec:tb}
\begin{figure}[ht]
\caption{\footnotesize Comparison of normal quantile-to-quantile plots relating to several numerical representations of the logarithmic bids. Left to right: mode specific normalization, minmax normalization, and standardization. Here, mode-specific normalization was applied to the bids in hope that it would provide a precise representation of the target, concerning the incumbent regression task.}
\centering
\includegraphics[scale=0.5]{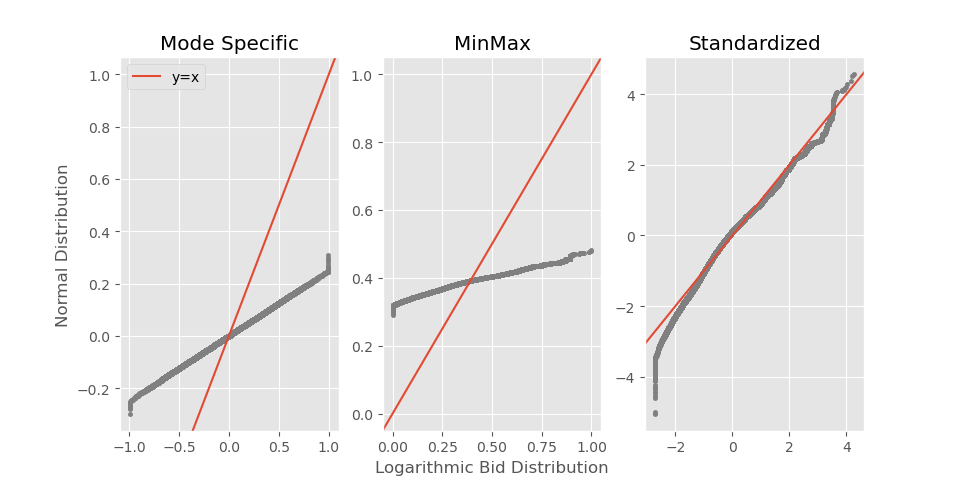}
\label{fig:targetqqplot}
\end{figure}

Recall that $B$ must represent all the firms at once by learning a bidding function based on auction features, which are synthesized with $A$. Ultimately, the best model from the two trained for $A$ will be selected. However, for now, this information is irrelevant, as only inputs sampled from the original data need to be considered to train an approximator $B:\mathbf{c}\xrightarrow{p(b|\mathbf{c}; \beta)}\hat{\theta}$. As shown by Figure \ref{fig:targetqqplot}, the standardized logarithmic bid distribution is Gaussian. Accordingly, the assumption of log-normality for the conditionals can be made, meaning that $\theta = (\mu, \sigma^{2})$, where $\mu$ and $\sigma^{2}$ are the first two moments of the Gaussian density.

In accordance with the pipeline delineated earlier, the input space of the bid generator $B$ must correspond to the output space of the auction generator $A$. The challenge arises due to the necessity of generating bids separately yet not independently from their conditional vector. Given that $\theta$ is a bi-dimensional vector, the task is defined as a multi-output regression. Since a multi-output linear model only represents the fitting of independent models and disregards the statistical dependence between each model's parameter set, a nonlinear approximator is required to model $B$.

A neural network, or more specifically a multi-layer perceptron (MLP), is the most comprehensive approximator capable of performing multi-target regression while maintaining the statistical dependence among its parameters \cite{Goodfellow2016b}. Taking an input vector $x$ and a target $y$, an MLP can represent the function $f(x; \gamma)$ by producing parameters for a distribution over $y$ instead of directly predicting $y$ \cite{Williams1996, Neuneier1994, Schioler1997}. This is applicable when $\gamma$ is optimized with respect to $-log P(y|x)$. Considering an MLP for $B$, its output is interpreted as a bi-dimensional vector containing the predicted first two moments of a conditional distribution $p(b|\mathbf{c})$. This model will be referred to as \textit{BidNet}.

To ensure the stability of BidNet (i.e., systematic convergence), it was trained on different folds using a cross-validation procedure. The methodology involves dividing the training set into $K$ folds (in this case, $K=5$) and utilizing $K-1$ folds to optimize the parameters with respect to the negative log-likelihood (NLL). Subsequently, the aggregated loss is computed on the remaining fold. After each pass over the training set, the validation set (the remaining fold) is employed to assess the objective NLL, which forms the basis for an early stopping rule. This strategy serves as a regularization method to prevent overfitting. The process is repeated until all $K$ folds have been used as validation sets. Although various early stopping designs have been identified by \cite{Prechelt2012}, a customized one was employed to best suit the requirements of this study. The pseudocode related to the training of BidNet with cross-validation is presented in Algorithm \ref{alg:bidnet} in Appendix C.

\subsubsection{Reinforcement Learning versus Supervised Learning for Bid Generation}
\label{sec:RLvsSL}

BidNet leverages supervised learning (SL) to predict bid distribution parameters, contrasting with the potential use of reinforcement learning (RL) for direct bid generation. RL is a machine learning approach where agents learn to make decisions by performing actions in an environment to achieve some notion of cumulative reward. It operates by utilizing feedback from its own actions and experiences in a dynamic environment.

Despite RL's prowess in decision-making and strategy development across various domains \cite{mnih2015, szepesvari2010, Yu2021}, it introduces uncertainties in synthetic data generation. RL's dependence on model-based rewards or realistic reward signals complicates the generation of realistic synthetic bids due to assumptions about agent rationality and objectives. In other words, the emergence of realistic behavior is not assured.

Furthermore, RL, particularly in multi-agent scenarios, faces computational stability and convergence challenges, unlike SL's reliability with labeled data. When high-quality labels are available, SL proves more effective, as in our case with BidNet, which accurately replicates bid signals. RL's exploration of strategy space and the complexity of achieving rational and convergent behavior add layers of difficulty, whereas SL offers a more stable and realistic approach to data generation \cite{mguni2018, riley2021, jin2009}.

The intricacies of modeling individual firms with RL, compounded by the lack of detailed firm features, would demand extensive engineering and potentially introduce biases. Thus, for purposes of replicating public procurement data and avoiding the risks of unrealistic coordination or collusion, SL and BidNet present a preferable methodology. BidNet efficiently utilizes bid signals for modeling, sidestepping the complexities and risks associated with RL, including the uncontrolled emergence of collusion patterns \cite{ezrachi2020, Klein2018, Ittoo2017}. This approach aligns with the goal of generating grounded synthetic data for analysis, offering a stable alternative for prediction and synthetic realism without the need for detailed assumptions about bidding strategies.

However, RL should not be overlooked, especially given its successful application in representing artificial bidders within electricity market environments \cite{viehmann2021, Ye2019, Rashedi2016}. In fact, as discussed in Section \ref{sec:FurtherResearch}, our method is not in competition with RL; rather, both approaches can be utilized synergistically, to the benefit of auction design research.

\subsection{Sampling Synthetic Auction Instances}
\label{subsec:sampling}
The GAN-based model relies on its generator to sample synthetic features from noise. However, following the principle of training-by-sampling, the user must also create a conditional vector or a batch of conditional vectors alongside the latent space $\mathbf{z}$. To achieve this, one must apply steps 3 to 6 described in Algorithm \ref{alg:ctwgan} and then feed the trained generator with the noise and the conditional vector, as illustrated in Figure \ref{fig:ctwgan}. A key advantage of this approach is that one can manually specify a conditional vector by disregarding steps 3 to 5 in Algorithm \ref{alg:ctwgan}, enabling the replication of the signal of interest. The tabular VAE also samples through latent noise that can be created from a standard Gaussian distribution.

A synthetic instance of public procurement comprises a vector of auction features $\tilde{\mathbf{c}}$ and its associated array of bids. It is important to note that the process of bid generation occurs in two steps because the number of bids to be generated per auction varies. The variable "number of bidders" encodes this variability and has been included in the joint distribution of auction characteristics; consequently, it is included in the output of $A$. Since BidNet predicts the first moments of a Gaussian distribution, it is straightforward to sample $nb$ bids from a random generator. In fact, when given synthetic inputs originating from $A$, BidNet provides $\tilde{\theta} = (\tilde{\mu}, \tilde{\sigma}^2)$, which are the parameters for the conditionals $p(b|\mathbf{c};\theta)$. Synthetic bids are then drawn according to $\tilde{b}\sim\mathcal{N}(\tilde{\theta})$.

To maintain consistency with the notation introduced in Figure \ref{fig:architecture}, we distinguish $\tilde{b}$ from $\hat{b}$, the latter representing the predicted bids originating from $\mathcal{N}(\hat{\theta})$, where $\hat{\theta}$ is the output of BidNet when provided with an original sample from the test set $\mathbf{c}_{test}$. It is important to note that the predicted bids $\hat{b}$ should be used for validation purposes only.

\section{Validation}
\label{sec:4}

We assess the faithfulness of the synthetic auction features, and by extension the performances of our synthesizers, by employing an inception score. The concept of inception scoring is based on the principle that a classifier or a regressor that has been trained successfully using synthetic data should perform well when tested on real-world instances. This principle underlies the expectation that the characteristics of a predictor's output are primarily determined by its input, given a fixed set of parameters. 

Inception scoring is a well-established method for validating the output of GANs and has been widely used for this purpose in the literature. The use of inception scoring as a validation metric for GANs was first proposed by \cite{Salimans2016} and has since been adopted and expanded upon by numerous researchers upon investigation on its usefulness as a metric for evaluating the quality of generated outputs \cite{Lucic2018}. 

In our case, we perform inception scoring by utilizing the binary variable \textit{municipality}, one of the auction characteristics, as a target in the following binary classification problem:

\[
f(\mathbf{c}_{-mun}) = p(mun),
\]

where $\mathbf{c}_{-mun}$ is the one-hot encoded set of auction features that excludes \textit{municipality}. In other terms, we define the classifier $f:\xrightarrow{p(mun|\mathbf{c}_{-mun})}[0,1]$. If the synthetic data is realistic, that is, if the synthesizer being evaluated managed to efficiently approximate the targeted joint distribution, then we expect the overall accuracy of $f(\mathbf{c}_{-mun})$ and $f(\tilde{\mathbf{c}}_{-mun})$ to be similar; provided $f$ has been successfully trained.

\begin{table}[!ht]
\centering
\caption{\footnotesize The classification accuracy for each class (Recall), as well as the average F1-score are reported. The classifiers have been trained two times each, using 100,000 training examples generated with the GANs and VAE-based method. Both synthesizers have been previously trained over 200 epochs. The numbers in the parenthesis indicate the performance gap (in percentage) between scores achieved on synthetic and real test-beds.}
	\subcaption*{(a) Evaluation metrics of models trained on synthetic data generated by the GAN-based method.}
	\begin{tabular}{c c c c c}
	\toprule
	     \textbf{Test-bed} & \textbf{Model} & \textbf{Recall (0)} & \textbf{Recall (1)} & \textbf{F1-score}\\
		\hline
		\multirow{3}{*}{Synthetic}
		        & Decision Tree & 0.89 & 0.85 & 0.87 \\
        	    & k-NN & 0.80 & 0.84 & 0.83 \\
		        & CMLP & 0.78 & 0.78 & 0.78 \\
	    \bottomrule
		\multirow{3}{*}{Real}
		        & Decision Tree & 0.86 (-0.03) & 0.48 (-0.32) & 0.67 (-0.23) \\
		        & k-NN & 0.70 (-0.12) & 0.78 (-0.07) & 0.74 (-0.11) \\
		        & CMLP & \textbf{0.81} (\textbf{+0.04}) & \textbf{0.79} (\textbf{+0.01}) & \textbf{0.80} (\textbf{+0.03}) \\
    	\bottomrule
	\end{tabular}
	\bigskip
	\subcaption*{(b) Evaluation metrics of models trained on synthetic data generated by the VAE-based method.}
	\begin{tabular}{c c c c c}
	\toprule
	     \textbf{Test-bed} & \textbf{Model} & \textbf{Recall (0)} & \textbf{Recall (1)} & \textbf{F1-score}\\
		\hline
		\multirow{3}{*}{Synthetic}
		        & Decision Tree & 1.00 & 1.00 & 1.00 \\		
		        & k-NN & 0.98 & 0.99 & 0.98 \\
	        	& CMLP & 0.98 & 0.98 & 0.98 \\
	    \bottomrule
		\multirow{3}{*}{Real}
		        & Decision Tree & 0.98 (-0.02) & 0.00 (-1.00) & 0.36 (-0.54)\\
		        & k-NN & 0.96 (-0.02) & 0.17 (-0.83) & 0.51 (-0.48)\\
	        	& CMLP &  0.46 (-0.53) & 0.99 (+0.01) & 0.69 (-0.30) \\
    	\bottomrule
	\end{tabular}
\label{tbl:val}
\end{table} 

Three models were utilized to represent the classifier $f$: a \textit{decision tree}, \textit{k-nearest neighbors} (k-NN) and a \textit{multi-layer perceptron}. Here, we will name our MLP classifier as CMLP, where C stands for classification, in order to distinguish this classifier from the MLP we used earlier to predicts the bids. Evaluation metrics for the three classifiers when trained on synthetic examples generated by the GAN-based and VAE-based methods are reported by Table \ref{tbl:val}. Based on the results, we can draw a clear conclusion: the GAN-based model succeeded in synthesizing reliable and realistic data, while the VAE-based model did not. Indeed, the CMLP achieved an overall classification F1-score on the real test-bed 3\% above what was attained on the synthetic test-bed. 

In contrast, when examining the VAE-based model, classifiers display relatively poor F1-scores, ranging from 36\% (Decision Tree) to 69\% (MLP). An F1-score below or around 50\% suggests that the classifier was not better than a random choice algorithm. In addition, the decision tree and the k-NN both assigned almost all instances of the real test-bed to the class 0. The MLP performed decently in recognizing some structure in the data generated by the VAE-based model. 

This gap in performance between the GAN and VAE-based methods explains why we indulged in the specification of two methods in the first place. It should also be noted that the results in this kind of experiment may vary according to the hyperparameter settings, and in theory, the tabular VAE should be able to reach a similar level of effectiveness with some fine tuning. For instance, the unrealistically high scores achieved on the TVAE synthetic test-bed can be a sign of over-training. The point to be made here, is that the GAN-based model provides a reliable way to perform the task at hand without having to worry too much about hyperparameter tuning. Details about our hyperparameter tuning are to be found in Appendix B.

Now, we need to evaluate the efficiency of BidNet, and by extension the reliability of the synthetic bids. An account of BidNet's relative efficiency is given by comparing it to a \textit{regression tree} model and a \textit{multi-target support vector regressor} (MSVR). The three regressors have been evaluated using the same metric, the negative log-likelihood (NLL), and their performances over five folds are reported in Table \ref{tbl:bidnet}. BidNet is, on average, the best model. Note that both the MSVR (Multi-output Support Vector Regression) and the regression tree do not inherit the topological flexibility of a neural network, such as BidNet, which allows for the nuanced modeling of complex non-linear relationships within data. In fact, a neural network can be trained to predict the parameters $(\mu, \sigma^2)$ of conditional bid distributions directly using a Negative Log-Likelihood (NLL) loss and outputting these parameters through a network structure designed to capture and represent hierarchical features. Conversely, the MSVR and the regression tree are constrained to a more direct form of prediction. Namely, they have been trained to explicitly predict these parameters using the empirical bid means and variances as targets via a multi-output wrapper.

Nevertheless, since the NLL is a relative measure with values that can range from $-\infty$ to $+\infty$, a thorough investigation of  BidNet's outputs is needed in order to assert the reliability of the resulting synthetic bid distribution. To that end, we propose a procedure for evaluating the distance between the distributions of fake and real bids, $Dist(p(b) || p(\tilde{b}))$, using the distance between the real and predicted distributions $Dist(p(b) || p(\hat{b}))$ as an identity, and the distance between the predicted and fake distributions $Dist(p(\hat{b}) || p(\tilde{b}))$ as a control.

\begin{table}[ht]
\caption{\footnotesize The averages and standard deviations of the negative log-likelihood (NLL) over the five cross-validation folds are given for BidNet, a tree-based regressor and a multi-target support vector regressor (MSVR). The column \textit{best} reports the NLL  to the best model identified with each method.}
	\centering
	\begin{tabular}{l c c c}
	\toprule
	    & \multicolumn{3}{c}{\textbf{NLL}} \\
	    \textbf{Model} & mean & std & best  \\ 
		\hline
	    BidNet & \textbf{0.59} & 0.08 & \textbf{0.56}  \\
	    MSVR & 1.16 & 0.02 & 1.12  \\
    	Regression Tree & 92,999 & 885 & 91,757 \\
	\bottomrule
	\end{tabular}
\label{tbl:bidnet}
\end{table} 

This procedure is also called "double validation" because it provides another way to validate the output of the synthesizers. Indeed, the opportunity to use BidNet in order to construct an inception score naturally occurs since it has been trained on the real data and generates synthetic bids from fake auction features. The results displayed in Table \ref{tbl:doubleval} are coherent with those of the previous validation as the tabular VAE is at the origin of a noisy synthetic bid distribution. Meanwhile, $Dist(p(b) || p(\tilde{b}))$ and $Dist(p(\hat{b}) || p(\tilde{b}))$ are very close to each other in both cases and also close to $Dist(p(b) || p(\hat{b}))$ in the GAN-generated data case. BidNet is hence effective in preserving the statistical dependence between bids and their auction features and is a powerful approximator of $p(b|\mathbf{c})$.

\begin{table}[!ht]
\centering
\caption{\footnotesize The statistical distances have been measured with the \textit{Wasserstein} or \textit{Earth-Mover Distance} (EMD), and the \textit{quantile-to-quantile root mean squared error} (QQ-RMSE). A score of 0 indicates identical distributions.}
    \subcaption*{(a) Synthetic bids emanate from the real data.}
	\begin{tabular}{l c c}
	\toprule
	     & \textbf{QQ-RMSE} & \textbf{EMD} \\
		\hline \\
		$Dist(p(b) || p(\hat{b}))$ & 1.194 & 0.003 \\
	\bottomrule
	\end{tabular}
	\bigskip 
    \subcaption*{(b) Synthetic bids emanate from the data generated by the GAN-based method.}
	\begin{tabular}{l c c}
	\toprule
	     & \textbf{QQ-RMSE} & \textbf{EMD} \\
		\hline\\
		$Dist(p(b) || p(\tilde{b}))$ & 1.364 & 0.006  \\
        $Dist(p(\hat{b}) || p(\tilde{b}))$ & 1.360 & 0.004  \\
	\bottomrule
	\end{tabular}
	\bigskip 
    \subcaption*{(c) Synthetic bids emanate from the data generated by the VAE-based method.}
	\begin{tabular}{l c c}
	\toprule
	     & \textbf{QQ-RMSE}  & \textbf{EMD} \\
		\hline\\
		$Dist(p(b) || p(\tilde{b}))$ & 122.762 & 0.284  \\
        $Dist(p(\hat{b}) || p(\tilde{b}))$ & 122.800 & 0.286  \\
	\bottomrule
	\end{tabular}	
\label{tbl:doubleval}
\end{table} 

\section{Discussion}
\label{sec:5}

In this study, we have investigated the application of GANs and VAEs for crafting realistic auction simulations. We employed these data-driven methods to generate synthetic auction features and bids, aiming to provide a credible and useful tool for researchers and practitioners in the field of auction design and analysis. To ensure the reliability of the generated data, we evaluated the performances of the synthesizers by employing inception scores and assessing the faithfulness of the synthetic auction features. Additionally, we compared BidNet's efficiency to regression tree and multi-target support vector regressor models to further validate the newly generated synthetic data.

Our results demonstrated that while both GANs and VAEs are capable of generating realistic artificial data, the CTGAN outperformed the tabular VAE in our context. The CTGAN's ability to generate data according to hand-crafted conditional vectors made it particularly suitable for simulating auction scenarios in our study. This finding highlights the importance of selecting the most appropriate method for the specific context and requirements of a given task.

In conclusion, this study offers a valuable contribution to the field of auction simulations by demonstrating the potential of GANs and VAEs for generating realistic and reliable synthetic data. While our approach is not without its limitations, the findings provide a strong foundation for future research in this area, particularly in exploring the potential of CTGANs and other generative models for simulating complex auction scenarios and their applications in various economic contexts.

\subsection{Implications for Further Research in Economics and for Practitioners}
\label{sec:FurtherResearch}

The availability of high-quality synthetic data opens avenues for training more complex or data-intensive machine learning algorithms, increasingly prevalent in economics. These advanced algorithms often require vast datasets for optimal training, a need that realistic synthetic data can fulfill. The richness and variety of this synthetic data allow for a more robust training process, contributing to the development of more accurate and sophisticated economic models. For instance, combining our method with other machine learning techniques, like those used in \cite{Chu2022} to forecast GDP growth, could enable researchers to predict various economic indicators or outcomes based on synthetic data.

The relevance and utility of realistic synthetic data generation are expected to grow in tandem with the rising prominence of Reinforcement Learning in economic research. RL agents, inherently data-intensive, necessitate extensive datasets for training. For example, our method could be integrated with multi-agent systems, to simulate more complex interactions and behaviors in auction markets \cite{Lussange2021}, or the method could be extended to simulate other types of auctions and market mechanisms \cite{Zhou2019}. Realistic synthetic data provides the necessary foundation for creating environments that are not only rich in information but also grounded in reality. This is crucial for ensuring that RL agents operate on realistic assumptions, leading to quality decision-making across various economic contexts. The RL capacity to adapt and learn from dynamic environments makes it a powerful tool in economic research, where understanding and predicting complex market behaviors and trends is essential.

Specifically within the domain of auction research, RL has shown promise in understanding and optimizing auction mechanisms. In electricity markets, for example, RL can be used to model and predict market behaviors, assisting in the design of more efficient and effective auction systems \cite{Shafie-Khah2015, Tellidou2007, zhao2019}. RL's ability to learn and adapt to complex auction dynamics makes it a valuable tool in this field. Similarly, RL's application in auction design extends beyond electricity markets, offering insights into various auction formats and strategies. By training on realistic auction data, RL algorithms can help in devising auction mechanisms that are more efficient, transparent, and beneficial to all stakeholders involved. This aspect of RL in auction design highlights its potential in reshaping how auction markets are studied and optimized \cite{Kimbrough2005, Waltman2008}.

In Section \ref{sec:RLvsSL}, we explored the potential of RL for bid generation and the reasons it was ultimately not adopted. Crucially, our discussion emphasizes the complementary nature of RL with our methodology, enhancing the scope for auction design research in public procurement. Specifically, our auction generator can supply rich, synthetic data environments for RL-based studies of bidding strategies within specific industries or markets. This synergy allows public authorities to gain deep insights, such as forecasting costs influenced by auction parameters, partly determined by these authorities themselves, thereby refining decision-making processes in public procurement.

Finally, auction simulation represents a powerful tool for both public procurement authorities and bidding entities, offering significant advantages in understanding and improving auction dynamics. First, for public procurement authorities, these simulations serve as a crucial testing ground. By simulating various auction designs, authorities can gather valuable feedback, especially in the context of preventing collusion and algorithmic collusion. This proactive approach allows them to refine and optimize auction structures, ensuring a fair and competitive bidding environment. Second, firms or bidding entities can leverage these simulations to gain a strategic edge. By representing themselves or other bidders through artificial agents, they can explore and test different bidding strategies. This experimentation can lead to a deeper understanding of auction mechanics and potentially foster tacit cooperation among bidders, which could be both a strategic advantage and a regulatory challenge. In conclusion, auction simulations offer potential benefits to all involved parties; they may enhance the fairness and efficiency of the auction process, but also allow for strategic insights and innovations, making them a valuable tool in the modern auction landscape.

\subsection{Limitations of Deep Generative Modeling}
It is important to acknowledge the limitations of our approach. A primary limitation is that GANs and VAEs can only capture the dynamics of data structures through time based on past data. Consequently, simulations generated using data-driven methods like ours cannot incorporate novel shock scenarios. This limitation is not unique to our approach, as no model can accurately predict the future, and anticipating future shocks that substantially modify the data generation process of a joint distribution remains an ongoing challenge in economics in general.

\section*{Statements and Declarations}
\subsection*{Acknowledgement}
We express our sincere gratitude to the reviewers for their insightful feedback, which has significantly contributed to the enhancement and refinement of this manuscript.

\subsection*{Funding}
This work was supported by Meta Research following the authors application to the request for proposals on \href{https://research.facebook.com/research-awards/request-for-proposals-on-agent-based-user-interaction-simulation-to-find-and-fix-integrity-and-privacy-issues/#award-recipients}{agent-based user interaction simulation to find and fix integrity and privacy issues}.

\subsection*{Competing Interests}
The authors declare no competing interests.

\subsection*{Author Contributions}
All authors contributed to the study's conception and design. Material preparation, data collection, analysis were performed by Igor Sadoune. The first draft of the manuscript was written by Igor Sadoune and all authors commented on previous versions of the manuscript. All authors read and approved the final manuscript.

\clearpage 
\section*{Appendix A: SEAO Dataset}

Table \ref{tbl:datadescription} outlines the variables used in our study. It is important to note that geographical variables like \textit{countries} or \textit{states} were excluded because, although present in the raw data, only contracts from the province of Quebec are actually recorded in the subset we accessed.

\begin{table}[ht]
    \caption{Description of the variables used in the study.}
    \centering
    \begin{tabular}{l p{10cm}}
        \toprule
        \textbf{Variables} & \textbf{Description} \\
        \hline
        \textit{public contractor} & The public entity that creates and publishes the contract for an auction. \\
        \textit{municipality} & Indicates whether the contract is issued at the municipal level. \\
        \textit{sector} & The industry sector from which the contract is issued. \\
        \textit{subsector} & The specific activity sector of the contract. \\
        \textit{location} & Various locations in Quebec where the contract is executed. This variable also includes combinations of locations for contracts spanning multiple areas. \\
        \textit{unspsc} & A global classification system for goods and services. \\
        \textit{number of bidders} & The number of firms participating in an auction. \\
        \textit{post-auction expenses} & Indicates any additional expenses paid by the public contractor after the contract is awarded. \\
        \textit{firms} & Firms associated with each bid in an auction. \\
        \textit{bids} & Standardized log bids, originally in Canadian dollars (CAD) in the raw data. \\
        \bottomrule
    \end{tabular}
    \label{tbl:datadescription}
\end{table}

The cleaning of the raw data involved selecting relevant variables. Many columns in the raw data were excluded as they did not provide informative signals due to their nature (e.g., web links) or redundancy with other variables. The quality of the signals was another selection criterion. Columns like temporal variables or textual entries were omitted due to low quality, evidenced by inconsistencies, excessive missing values, and non-uniform entry formats. Despite centralized dataset management by SEAO, data inconsistencies and missing values are common due to manual updates by various public entities' administrators. These are significant limitations of this dataset.

The raw data, is available at \url{https://www.donneesquebec.ca/recherche/dataset/systeme-electronique-dappel-doffres-seao}, and the official PDF description file at \url{https://www.donneesquebec.ca/recherche/dataset/d23b2e02-085d-43e5-9e6e-e1d558ebfdd5/resource/af41596c-b07f-4664-82c8-577e1ef9a6f3/download/seao-specificationsxml-donneesouvertes-20171010.pdf}. Data for each year, and in some cases each month, must be fetched separately. The raw data, provided in XML format, needs conversion into a workable tabular array. We utilized the Python "xml" library to convert and save the data in pickle (.pkl) format. The code for processing the original XML files is available in the associated GitHub repository for this manuscript. Note that we also provide the cleaned and preprocessed pickle file.

\newpage
\section*{Appendix B: Methodology Overview}

Initially, the SEAO dataset underwent a thorough cleaning process. This involved handling missing values, removing irrelevant columns, and reformatting specific columns to ensure their consistency and reliability.

Following the data cleaning, preprocessing was conducted to transform the dataset and make it suitable for machine learning applications. Discrete variables were transformed using one-hot encoding techniques, while continuous bid values were standardized.

To generate synthetic data, two primary generative models were utilized: CTGAN and TVAE. The CTGAN model was trained using an array of hyperparameters, including distinct embedding dimensions, generator and discriminator dimensions, learning rates, and specific decay rates. Likewise, the TVAE, a variant that incorporates a variational autoencoder structure, was trained with specific parameters, including hidden and latent dimensions.

Once trained, these models were then employed to sample synthetic datasets, replicating the patterns and distributions seen in the original SEAO dataset.

Subsequent to the synthetic data generation, we introduced BidNet neural network model. This model was designed to predict bid values using both discrete and continuous inputs. For training efficiency, the model utilized cross-validation and early stopping methodologies. Several hyperparameters, including learning rate, batch size, and number of epochs, were tuned manually to enhance the model's performance. Alternatively, rigorous automated tuning procedures (e.g., Bayesian optimization) can be used, provided enough computational resources and time are available.

To assess the quality of the synthetic data produced by CTGAN and TVAE, a series of classifiers, including Decision Trees, k-Nearest Neighbors, and Neural Networks, were trained on both the real and synthetic datasets. Performance metrics from these classifiers provided insights into the fidelity and utility of the synthetic data.

Finally, BidNet model's performance was critically evaluated using various metrics. These metrics, namely the Root Mean Square Error (RMSE), Jensen-Shannon distance (JS), and Wasserstein distance (WS), compared the synthetic and real bids, giving a comprehensive understanding of the model's accuracy and effectiveness.

Throughout the entire process, special attention was given to reproducibility. Foundational functionalities ensured consistent random states, allowing for deterministic behavior across runs. Additionally, capabilities were established to save intermediate results, trained models, and to manage computation across various devices, whether CPU or GPU.

\newpage
\section*{Appendix C: Algorithms}

\begin{algorithm}
\caption{Training GANs-based auction features generator}
    \begin{algorithmic}[1]
        \State $D_{train} \gets$ Initialize training set
        \While {$C(fake)>threshold$} \Comment{The critic can be optimized until $C(fake)$ is near $0$. } 
            \State Randomly select a discrete variable $c$ with equal probability
            \State Compute the probability mass function (PMF) of $c$
            \State Randomly select a state $i^*$ inherent to $c$ according its PMF
            \State Create the conditional vector $cond$ so that $\sum_i cond(i) = 1$ and $cond(i^*)=1$
            \For {$batch \in \{1,\dots,N_{batches}\}$} \Comment{Gradient descent with mini-batch}
                \State $real \gets d(c_{i^*}=1)\sim D_{train}$ \Comment{Sample batch of real examples respecting the constraint}
                \State $z \sim \mathcal{N}(0,1)$ \Comment{Sample noise}
                \State $fake \gets \tilde{d} \sim G(z)$ \Comment{Sample fake examples}
                \State $real \gets [real]\times 10$ \Comment{Stack input 10 times for Pac configuration}
                \State $fake \gets [fake] \times 10$
                \State $L^j \gets \big( C(fake_j) - C(real_j)\big) + CE(\tilde{c}, cond)$
                \State $L^{batch} \gets L^{batch} + \lambda(||\nabla L^{batch} ||_{2} - 1)^2$ \Comment{Apply gradient penalty}
                \State $w_{crit} \gets w_{crit} + Adam(\nabla_{w_{crit}}\frac{1}{m}\sum_i^m L^{batch}(i))$ \Comment{Updating $C$ with Adam}
                \If {$batch \bmod k = 0$} \Comment{Synchronicity, depends on $k$}
                    \State $w_{gen} \gets w_{gen} + Adam(\nabla_{w_{gen}}\frac{1}{m}\sum_i^m -C(G(z)))$ \Comment{Updating $G$ with Adam}
    				\EndIf
    			\EndFor
    		\EndWhile
    \end{algorithmic}
\label{alg:ctwgan}
\end{algorithm} 

\begin{algorithm}
\caption{Training tabular VAE for auction features}
    \begin{algorithmic}[1]
        \State $D_{train} \gets$ Initialize training set
        \For {$epoch\in N_{steps}$}
            \For {$batch \in \{1,\dots,N_{batches}\}$} \Comment{Gradient descent with mini-batch}
                \State $real \sim D_{train}$ \Comment{Sample batch of real examples}
                \State $(\mu, \sigma^2) \sim Enc(real)$
                \State $z \sim \mathcal{N}(\mu,\sigma^2)$ \Comment{Sample latent input}
                \State $fake \sim Dec(z)$ \Comment{Sample fake examples}
                \State $L^j \gets CE(\tilde{ c}_j - \argmax(c)) + (2\sigma^2)^{-1}(x_j - tanh(\tilde{x}_j))^2  + KL(\mathcal{N}(\mu_j, \sigma_j^2), \mathcal{N}(0,1))$
                \State $w \gets w + Adam(\nabla_{w}\frac{1}{m}\sum_i^m L^{batch}(i))$ \Comment{Updating parameters with Adam}
    			\EndFor
    		\EndFor
    \end{algorithmic}
\label{alg:tvae}
\end{algorithm} 

\begin{algorithm}
\caption{K-folds cross-validation BidNet training procedure}
    \begin{algorithmic}[1]
        \State $D \gets \{D_1,\dots,D_K\}$ \Comment{Initialize K-folds}
        \State $loss^{*}\gets \infty$ \Comment{Initialize best model}
        \For {$fold \in D$}
            \State $reset(w_{BidNet})$ \Comment{Reset parameters before entering each new fold}
            \State $D_{val}\gets D(fold)$, $D_{train}\gets D(-fold)$
            \While {has not converged}
                \For {$batch \in \{1,\dots,N_{batches}\}$} \Comment{Gradient descent with mini-batch}
                    \State $d \sim D_{train}$ \Comment{Sample batch of real examples}
                    \State $\hat{\theta}\gets BidNet(d)$
                    \State $L^{train}\gets m^{-1}\sum_i NLL(\hat{\theta})_i$ \Comment{compute NLL on training batch}
                    \State $w \gets w + Adam(\nabla L^{train})$ \Comment{Update BidNet}
                \EndFor
                \State $converged \gets ES(L^{val})$ \Comment{Early stopping}
                \State $L^{val}\gets n^{-1} \sum_j NLL(BidNet(D_{val}))_j$ \Comment{compute NLL on validation fold}
                \If {$L^{val}<loss^{*}$}
                    \State $loss^{*} \gets L^{val}$
                    \State save model
                \EndIf
			\EndWhile
		\EndFor
    \end{algorithmic}
\label{alg:bidnet}
\end{algorithm} 

\begin{algorithm}
\caption{Synthetic bid validation / Double validation}
    \begin{algorithmic}[1]
        \State $\beta \gets \beta^*$ \Comment{load best set of parameters for BidNet}
        \State $\alpha \gets \alpha^*$ \Comment{load optimized set of parameters for synthesizer}
        \State $\tilde{\mathbf{c}}\sim A_{\alpha^*}(\mathbf{z})$\Comment{sample synthetic examples from the trained synthesizer}
        \State $\mathbf{c} \sim D_{test}$\Comment{sample a test-set of real instances}
        \State $\hat{b} \sim B_{\beta^*}(\mathbf{c})$\Comment{sample predicted bids from the test-set of real instances using BidNet}
        \State $\tilde{b} \sim B_{\beta^*}(\tilde{\mathbf{c}})$\Comment{sample fake bids with the synthetic data emanating from the synthesizer}
        \State $Dist(p(b) || p(\tilde{b}))$\Comment{compute the statistical distance between the fake and real distributions of bids}
        \State $Dist(p(b) || p(\hat{b}))$\Comment{compute the statistical distance between the predicted and real distributions of bids}
        \State $Dist(p(\hat{b}) || p(\tilde{b}))$\Comment{compute the statistical distance between the predicted and fake distributions of bids}
    \end{algorithmic}
\label{alg:doubleval}
\end{algorithm} 

\clearpage 

\addcontentsline{toc}{section}{References}
\bibliographystyle{utphys}
\bibliography{library}

\end{document}